\documentstyle[aps,prc,preprint,epsf,tighten,twoside]{revtex}
\begin{document}
\title{Heavy quarkonia in the instantaneous Bethe-Salpeter model}
\author{J.Resag, C.R.M\"unz}
\address{Institut f\"ur Theoretische Kernphysik,\\
         Universit\"at Bonn, Nussallee 14-16, 53115 Bonn, Germany}
\date{\today}
\maketitle

\begin{abstract}
  The heavy quarkonia (Charmonium \(c\bar{c}\) and Bottomonium
  \(b\bar{b}\)) are investigated in the framework of the instantaneous
  BS-equation (Salpeter equation).  We parametrize confinement
  alternatively by a linearly rising scalar or a vector interaction
  kernel and take into account the one-gluon-exchange (OGE)
  interaction in the instantaneous approximation. Mass spectra as well
  as leptonic, two-photon, E1 and M1 decay widths are calculated.  Our
  results show that a reasonable description of the experimental data
  can be obtained with both spin structures for the confining kernel.
  The relativistic treatment leads to an improved description compared
  to nonrelativistic results for the two-photon width of the
  \(\eta_c\) and to some extent for the E1-transition widths.
  However, characteristic deviations indicate that within a
  relativistic framework confinement is not described adequately by a
  potential.
\end{abstract} \pacs{}

\narrowtext

\section{Introduction}
In the past the heavy quarkonia have usually been investigated in the
framework of the nonrelativistic quark model (see e.g. \cite{Bey,LSG}
and references therein). Because of the large mass of the \(c\) or
\(b\) quark the nonrelativistic treatment of the bound state problem
is expected to be a good first approximation. However in charmonium
one still finds typical velocities of \(v/c \approx 0.4\) (see e.g.
ref.\cite{HaSu}), so that relativistic effects should become important
especially for electroweak decay properties, as has been shown in
ref.\cite{Bey}.

Relativistic calculations for the heavy quarkonia have been reported
e.g.  by Tiemeijer and Tjon \cite{tjo} who compare various
quasipotential approximations to the BS-equation, by Gara et.al.
\cite{Ga} within the framework of the reduced Salpeter equation, and
by Murota \cite{Mur} who uses the (full) Salpeter equation \cite{Sa}.
Unfortunately, these authors only give the mass spectra and do not
calculate any decay widths, which should be most sensitive to
relativistic effects.

In the present contribution we obtain the mass spectra as well as the
leptonic, two-photon, E1 and M1 decay widths in the framework of the
(full) Salpeter equation. We parametrize confinement by a linearly
rising scalar or a vector interaction kernel and take into account the
one-gluon-exchange (OGE) interaction in the instantaneous
approximation. The Salpeter equation is then solved numerically
according to the treatment outlined in ref.\cite{RMMP}.  The
calculation of the decay widths is performed in the Mandelstam
formalism \cite{Ma}.

The paper is organized as follows: In Sec.\ref{hI} we give the
explicit form of the interaction kernel and briefly review the
formalism for the calculation of the decay widths. The model
parameters and results are discussed in Sec.\ref{hII}, and we give
some concluding remarks in Sec.\ref{hIII}.

\section{The Model}  \label{hI}
\subsection{The Bethe-Salpeter kernel}
For an instantaneous BS-kernel and free propagators with effective
quark masses \(m_1\) and \(m_2\) one can perform the \(p^0\) integrals
in the BS-equation in the rest frame of the bound state with mass
\(M\) and thus arrives at the (full) Salpeter equation
\begin{eqnarray}
\Phi(\vec{p}) &=&
\int \!\!\frac{d^3p'}{(2\pi)^3}\,
\frac{\Lambda^-_1(\vec{p})\,\gamma^0\,
[(V(\vec{p},\vec{p}\,')\,\Phi(\vec{p}\,')]
\,\gamma^0\,\Lambda^+_2(-\vec{p})}
{M+\omega_1+\omega_2}
 \nonumber \\
 &-&
\int \!\!\frac{d^3p'}{(2\pi)^3}\,
\frac{\Lambda^+_1(\vec{p})\,\gamma^0\,
[(V(\vec{p},\vec{p}\,')\,\Phi(\vec{p}\,')]
\,\gamma^0\,\Lambda^-_2(-\vec{p})}
{M-\omega_1-\omega_2}
 \label{9}
\end{eqnarray}
with \(\omega_i=\sqrt{\vec{p}\,^2+m_i^2}\) and the projection
operators \( \Lambda^{\pm}_i(\vec{p}) = (\omega_i \pm
H_i(\vec{p}))/(2\omega_i) \) on positive and negative energies, where
\(H_i(\vec{p})=\gamma^0(\vec{\gamma}\vec{p}+m_i)\) is the standard
Dirac hamiltonian (for the notation we refer to refs.\cite{RMMP,Mue}).

The confinement plus OGE interaction kernel applied in the present
work reads
\begin{equation}
\left[V(\vec{p},\vec{p}\,')\,\Phi(\vec{p}\,')\right] =
\left[V_C(\vec{p},\vec{p}\,')\,\Phi(\vec{p}\,')\right] +
\left[V_G(\vec{p},\vec{p}\,')\,\Phi(\vec{p}\,')\right]
\end{equation}
where the scalar or vector confining part is given by
\begin{eqnarray}
\left[V_C^S(\vec{p},\vec{p}\,')\,\Phi(\vec{p}\,')\right]
&=& \;\;\;{\cal V}_C((\vec{p}-\vec{p}\,')^2)\,\Phi(\vec{p}\,')
\;\;\;\;\mbox{or}
\label{ccscalar} \\
\left[V_C^V(\vec{p},\vec{p}\,')\,\Phi(\vec{p}\,')\right] &=& -{\cal
    V}_C((\vec{p}-\vec{p}\,')^2)\,\gamma^0\,\Phi(\vec{p}\,')\,\gamma^0
\label{ccvector}
\end{eqnarray}
respectively.
Here \({\cal V}_C\) is a scalar function with the fourier
transform \({\cal V}_C^F(r) = a_c+b_c r\).

For the OGE kernel \(V_G\) we have to note that it is not possible to
formulate this term in a gauge-invariant way, since for a
gauge-invariant kernel it is essential to take into account crossed
gluon diagrams. However, for such diagrams the instantaneous
approximation cannot be applied in a straightforward way. Furthermore,
also in a noninstantaneous treatment the incorporation of crossed
diagrams is technically very difficult, so that it would be very hard
to go beyond the gauge-dependent ladder approximation.

In view of the instantaneous treatment of the OGE the natural gauge
for the gluon propagator is the Coulomb gauge, which will be applied
in the following. The advantage of this gauge is the fact that the
gluon propagator given by
\begin{equation}
\gamma^{\mu}\,D_{\mu\nu}(q)\,\gamma^{\nu}
= 4\pi\,\left( \frac{\gamma^0\gamma^0}{\vec{q}^{\,2}}
+ \frac{\vec{\gamma}\vec{\gamma}
     - (\vec{\gamma}\hat{q})(\vec{\gamma}\hat{q})}
       {q^2+i\epsilon} \right)
\end{equation}
with \(\hat{q}=\vec{q}/|\vec{q}|\)
is already instantaneous in its component \(D_{00}(q)\).
In the instantaneous approximation we substitute \(q^2\) by
\(-\vec{q}^{\,2}\). The OGE kernel
then reads \cite{tjo,Mur}
\begin{equation}
\left[V_G^C(\vec{p},\vec{p}\,')\,\Phi(\vec{p}\,')\right]
= {\cal V}_G((\vec{p}-\vec{p}\,')^2) \,\left[
\gamma^0 \Phi(\vec{p}\,') \,\gamma^0
-\frac{1}{2}\,\left(
\vec{\gamma} \Phi(\vec{p}\,')\, \vec{\gamma} +
(\vec{\gamma}\hat{x}) \Phi(\vec{p}\,')\, (\vec{\gamma}\hat{x})\,
\right)\,\right] \label{Couleq}
\end{equation}
with
\begin{equation}
{\cal V}_G(\vec{q}^{\,2}) =
4\pi\,\frac{4}{3}\,\frac{\alpha_s(\vec{q}^{\,2})}{\vec{q}^{\,2}}
\end{equation}
We don't specify the operator \(\hat{x}=\vec{x}/|\vec{x}|\)
explicitely in momentum space since the corresponding matrix elements
are evaluated in coordinate space.  In analogy to the treatment of the
confinement matrix elements in ref.\cite{Mue} also the matrix elements
of \({\cal V}_G(\vec{q}^{\,2})\) are evaluated in coordinate space.
For the numerical calculation we will therefore obtain an analytic
expression for the Fourier transformed OGE potential \({\cal
  V}_G^F(r)\) in the following.

In QCD the running coupling constant for \(Q^2=-q^2 \gg
\Lambda_{QCD}\) is given by \cite{PDG}
\begin{equation}
\alpha_s^{run}(Q^2) =
\frac{A}{\ln(Q^2/\Lambda_{QCD}^2)}
\,\left( 1 - B \,\frac{\ln\,(\ln(Q^2/\Lambda_{QCD}^2))}
       {\ln(Q^2/\Lambda_{QCD}^2)} \right)
 + \; \ldots \label{alfas}
\end{equation}
with
\begin{equation}
A=\frac{12\,\pi}{33-2\,n_f} \;\;,\;\;
B=\frac{6\,(153-19\,n_f)}{(33-2\,n_f)^2}
\end{equation}
where in the instantaneous approximation we set \(Q^2=\vec{q}^{\,2}\).
We will assume that \(\alpha_s(\vec{q}^{\,2})\) behaves like
\(\alpha_s^{run}(\vec{q}^{\,2})\) for \(\vec{q}^{\,2} \gg
\Lambda_{QCD}^2\) and reaches a saturation value \(\alpha_{sat}\) for
\(\vec{q}^{\,2} \ll \Lambda_{QCD}^2\) with some smooth interpolation
in between.

The Fourier transformation of the OGE kernel can now be performed
analytically in the short and long distance region.  For \(r \gg
\Lambda_{QCD}^{-1}\) only small \(\vec{q}^{\,2}\) are important in the
Fourier integral and we can set
\(\alpha_s(\vec{q}^{\,2})=\alpha_{sat}\) so that
\begin{equation}
{\cal V}_G^F(r) = \frac{4}{3}\,\frac{\alpha_{sat}}{r}
\;\;\;\mbox{for}\;\; r \gg \Lambda_{QCD}^{-1}
\end{equation}
Analogously for \(r \ll \Lambda_{QCD}^{-1}\) we set
\(\alpha_s(\vec{q}^{\,2})=\alpha_s^{run}(\vec{q}^{\,2})\)
and obtain (see Appendix \ref{alfasapp})
\begin{eqnarray}
  {\cal V}_G^F(r) &\approx& \frac{4}{3}\,\frac{\alpha_s^{run}(r)}{r}
  \;\;\;\mbox{for}\;\; r \ll \Lambda_{QCD}^{-1} \;\;\;\mbox{with} \\
  \alpha_s^{run}(r) &=& \frac{A}{2\,\ln(e^{-\gamma}/a)}\,
  \left[1-B\,\frac{\ln\,(2\,\ln(1/a))}{2\,\ln(1/a)} \right]
    \label{arun}\\ && \mbox{with}\;\;a=\Lambda_{QCD}\,r \nonumber
\end{eqnarray}
where \(\gamma=0.577215\ldots\) is the Euler-Mascheroni constant.
An interpolation between these two limiting
cases is given by
\begin{equation}
  \alpha_s(r) = \frac{A} {2\,\ln \left(e^{-(\gamma+\mu a)}/a +
    e^{A/(2\alpha_{sat})} \right)} \,
  \left[1-B\,\frac{\ln\,(2\,\ln(e^{-\tilde{\mu}a}/a + e^{1/2}))}
    {2\,\ln(e^{-\mu a}/a + e^{B/2})} \right] \label{alfasr}
\end{equation}
(see ref.\cite{tjo} for the case \(B=0\)), where we set \(\mu=4\) and
\(\tilde{\mu}=20\) in order to obtain a smooth behaviour for
intermediate \(r\).

The Salpeter equation with \({\cal V}_G^F(r)=(-4/3)\,\alpha_s(r)/r\)
is well defined. This is in contrast to the corresponding
Schr\"odinger equation where the terms of order \(\vec{p}^{\,2}/m^2\)
like the spin-spin and spin-orbit interaction lead to a collaps of the
wavefunction into the origin, i.e. the Fermi-Breit hamiltonian is
unbound from below. This defect is usually cured by using first order
perturbation theory or by regularizing the \(1/r\) potential for small
\(r\).

\begin{figure}
  \vspace{0.5cm} \centering \leavevmode \epsfxsize=0.80\textwidth
  \epsffile{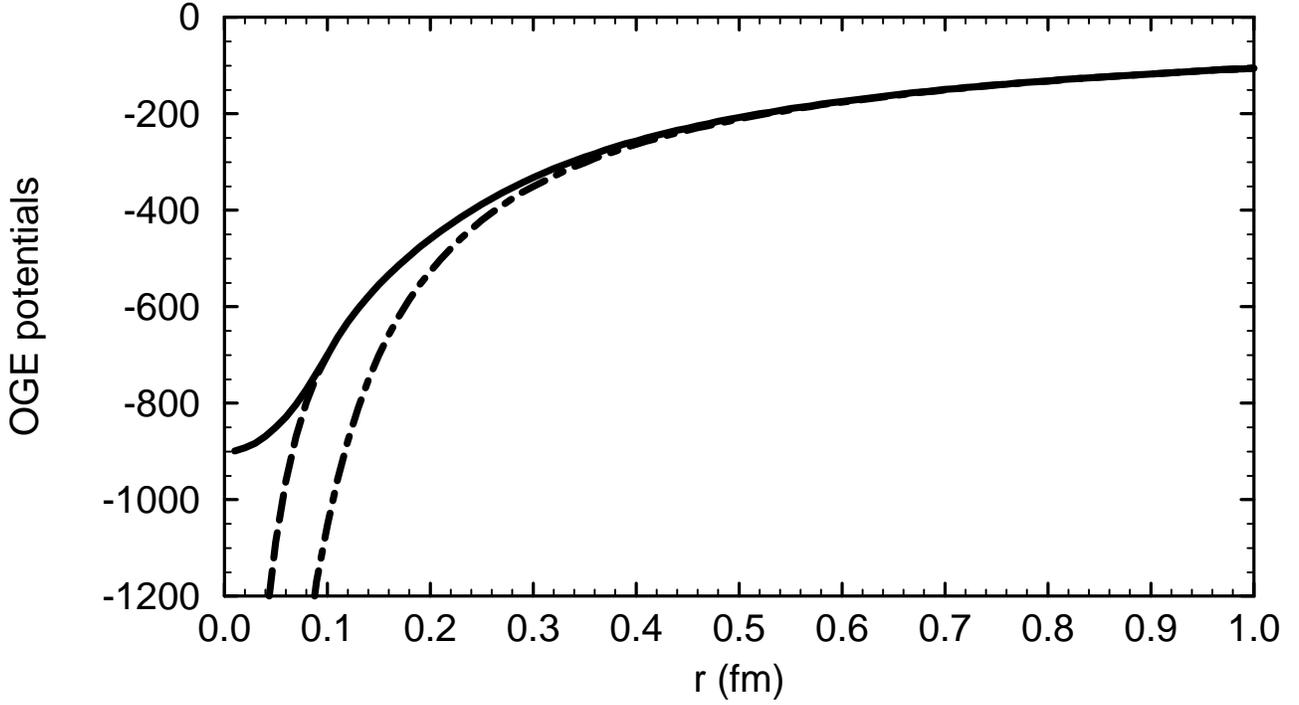} \vspace{1.5cm}
\caption{The regularized potential \({\cal V}_G^F(r)\) as given in
  eq.(\protect\ref{VGFr}) (solid curve) compared to the unregularized
  potential (dashed curve) and the potential
  \((-4/3)\,\alpha_{sat}/r\) (dashed-dotted curve) with
  \(\alpha_{sat}=0.4\).}
\label{VGfig}
\end{figure}

For the Salpeter equation this problem disappears due to the
relativistic treatment of the quark motion. However, most Salpeter
amplitudes are divergent for \(r \rightarrow 0\), as has been shown
explicitely by Murota \cite{Mur} for a fixed coupling constant. For a
running coupling constant this divergence is less pronounced, but
still present. The amplitudes are normalizable, but problems occur for
decay observables like the leptonic decay widths, which depend on the
value of the amplitudes at \(r \rightarrow 0\). The easiest way to
cure these divergencies is to regularize the OGE kernel for small
\(r\). We therefore will use the regularized potential
\begin{eqnarray}
{\cal V}_G^F(r) &=& -\frac{4}{3}\,\frac{\alpha_s(r)}{r}
\;\;\;\;\mbox{for}\;\;r > r_0 \nonumber \\
{\cal V}_G^F(r) &=& a_G\,r^2+b_G
\;\;\;\;\mbox{for}\;\;r \le r_0 \label{VGFr}
\end{eqnarray}
with \(a_G\) and \(b_G\) determined by the condition that \({\cal
  V}_G^F(r)\) and its first derivative are continuous functions.  The
dependence of \(\alpha_s(r)\) on \(\Lambda_{QCD}\) and \(n_f\) given
by eq.(\ref{alfasr}) is not strong and can be compensated for by
modifying \(\mu\) and \(\alpha_{sat}\). We will use
\(\Lambda_{QCD}=200\,MeV\) and \(n_f=3\) for our calculation.  A plot
of \({\cal V}_G^F(r)\) is shown in Fig.\ref{VGfig}.  The dependence of
the mass spectra on the regularization parameter \(r_0\) is very weak
so that the differences in the mass spectra calculated with the
regularized and unregularized potential are quite small. For our
further calculation we will take \(r_0=0.1 \,fm\).

\subsection{Calculation of decay widths}
The general prescription for the calculation of any current matrix
element between bound states has been given by Mandelstam \cite{Ma},
see e.g. \cite{Lu} for a textbook treatment.  The explicit formulas
for the leptonic and two-photon decay widths are given in
ref.\cite{RMMP}.  Since these transitions involve a non-hadronic final
state they can be calculated in the rest frame of the bound state
where also the amplitudes are determined.  The calculation of E1 and
M1 transitions, however, involves a boost of at least one of the meson
amplitudes. A covariant formulation of the Salpeter equation \cite{WM}
enables to treat this boost correctly, i.e.  we make the ansatz that
the BS-kernel \(K\) can be written covariantly as \(K(P,p,p') =
V(p_{\perp},\,p_{\perp}') \label{int}\) where \mbox{\( p_{\perp} =
  p-(Pp/P^2)\,P \)}, together with an analogous reformulation for the
spin structure of \(K\).  Explicitely eq.(\ref{Couleq}) can be
rewritten in a covariant form by replacing \(\vec{p} \rightarrow
p_{\perp}\) (and the same for \(\vec{x}\)), \(\gamma^0 \rightarrow
\gamma_{\mu}P^{\mu}/\sqrt{P^2}\) and \(\vec{\gamma} \rightarrow
-\gamma_{\mu}p_{\perp}^{\mu}/\sqrt{-p_{\perp}^2} \).

Since the details for the calculation of electromagnetic transitions
within the present framework have already been given in detail in
ref.\cite{MueForm}, we will only review the basic steps in the
following.

{}From the Bethe-Salpeter equation
\begin{equation}
\chi_P(p) = S^F_1(p_1)\,
\int \frac{d^4 p'}{(2\pi)^4}\,
[-i\,K(P,p,p')\,\chi_P(p')]\,
S^F_2(-p_2)
\end{equation}
with \(p_1=\eta_1P+p\), \(p_2=\eta_2P-p\) and \(\eta_1+\eta_2=1\) one
finds that the amputated BS amplitude or vertex function \(
\Gamma_P(p) := [S^F_1(p_1)]^{-1} \,\chi_P(p)\;[S^F_2(-p_2)]^{-1} \)
may be computed in the rest frame from the equal time amplitude \(
\Phi(\vec{p}\,) := \int\!  dp^0\,\chi_{(M,\vec{0})}(p^0,\vec{p}\,) \)
as
\begin{equation}
  \Gamma(\vec{p}\,) := \Gamma_{(M,\vec{0})}(p) =
  -i\! \int\!\! \frac{d^3p'}{(2\pi)^4}
  \left[ V(\vec{p},\vec{p}\,')\Phi(\vec{p}\,')\right]
\label{vert}
\end{equation}

\begin{figure}
  \centering \leavevmode \epsfxsize=0.70\textwidth
  \epsffile{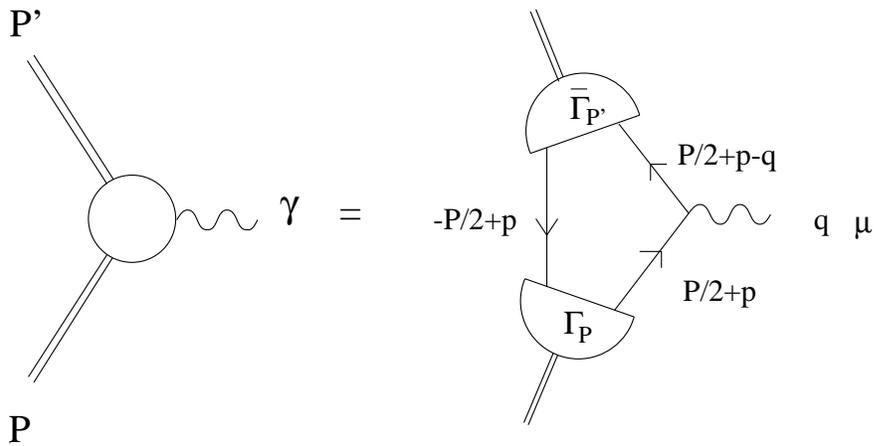} \vspace{1cm}
  \caption{The electromagnetic current $j^{(1)}_{\mu}$ coupling to the
    quark in lowest order calculated in the Mandelstam formalism from
    the BS vertex functions \protect{\(\Gamma_P,
      \bar{\Gamma}_{P'}\)}.}
  \label{Kgampt}
\end{figure}

Because of the covariant ansatz of the interaction kernel the
kinematical boost \(\Lambda_P\) with \(P=\Lambda_P\,(M,\vec{0})\)
gives the solution of the equation for any momentum \(\vec{P}\) of the
bound state, i.e.
\begin{equation}
  \chi_P(p) = \;
  S_{\Lambda_P}^{}\;\;\chi_{(M,\vec{0})}(\Lambda_P^{-1}p)\;\;
  S_{\Lambda_P}^{-1}.
\label{boo}
\end{equation}
(and \(\Gamma_P\) analogously).  The electromagnetic current between
two bound states may now be calculated from the BS amplitudes and a
kernel \(K^{(\gamma)}\) which is irreducible with respect to the
incoming and outgoing quark antiquark pair, i.e. it includes all
diagrams that may not be divided by just cutting the quark and the
antiquark line. In lowest order the matrix element of the
electromagnetic quark current taken between bound states with momenta
\(P\) and \(P'\) as shown in Fig.\ \ref{Kgampt} reads explicitely
\begin{eqnarray}
  \lefteqn{\left\langle\,P'\,\left|\,j_{\mu}^{(1)}(0)\,
  \right|\,P\,\right\rangle =} \\ &= & -
e_1\;\int\!\!\frac{d^4p}{(2\pi)^4}\;\; tr \; \Big\{
\bar{\Gamma}_{P'}(p-q/2)\;{S^F_1}(P/2+p-q)
\;\gamma_{\mu}\;{S^F_1}(P/2+p)\;\Gamma_{P}(p)\;{S^F_2}(-P/2+p)\Big\}
\nonumber
\end{eqnarray}
where $e_1$ is the charge of the quark and $q$ the momentum of the
photon. As in the BS-equation we will use \(S^F_i(p) = i/(p\!\!
/-m_i+i\epsilon)\).  The calculation of the current is performed in
the rest frame of the incoming particle, i.e. \(P=(M,\vec{0})\), the
results however are independent of this choice because of the formal
covariance. The $p^0$ integral picks up only the residues of the one
particle poles, the $\varphi_p$ dependence is trivial for decays in
z-direction and the resulting twodimensional integral in $|\vec{p}|$
and $\cos{\Theta_p}$ is calculated by Gaussian integration, compare
\cite{MueForm} for the details.

The electromagnetic decay width follows from the well known formula
for the decay rate with $\vec{q} = q\,\vec{e}_z$ the momentum,
$\lambda$ the polarization and $\varepsilon_\mu (\vec{q} ,\lambda)$
the polarization vector of the photon, i.e.
\begin{eqnarray}
 \lefteqn{ \Gamma(M\rightarrow M'\gamma) =}
\\
& = & \frac{1}{4\pi}\;\frac{k}{M^2}\; \frac{1}{2J+1}\, \sum_{M_J\,
  M_{J'}}\, \left|\,\varepsilon_\mu (\vec{q} ,\lambda=+1) \;
\left\langle\,P'\, J'\,M_{J'}\,\right|\,j^\mu(0)\,
  \left|\,P\,J\,M_J\,\right\rangle\right|^2 \nonumber
\end{eqnarray}

\section{Results and discussion} \label{hII}
\subsection{The model parameters}
We investigate two different models of the
confinement kernel: 1) a scalar \( 1\otimes 1\)-
and 2) a vector \(\gamma^0\otimes\gamma^0\)-structure.

The parameters used are the charm and bottom quark masses \(m_c\) and
\(m_b\), the offset \(a_c\) and slope \(b_c\) of the confinement
interaction and the saturation value \(\alpha_{sat}\) for
\(\alpha_s(r)\) in eq.(\ref{alfasr}).  These five parameters have been
adjusted to the mass spectra by minimizing a \(\chi^2\) that
incorporates all known charmonium and bottomonium ground states and
first excited states.  The resulting parameter sets are given in
Tab.\ref{ccbbpar} for the scalar (S) and the vector (V) confinement.

The main difference between the two parameter sets is given by the
larger value of \(\alpha_{sat}\) for the scalar confinement. This can
be easily understood from the nonrelativistic picture where the
spin-orbit force coming from the scalar confinement counteracts the
OGE spin-orbit force, whereas for the vector confinement both
spin-orbit forces affect the mass spectra in the same way. Therefore,
in order to compensate the reduced spin-orbit splitting of the
\(\chi\)-states in the scalar confining case, the strength of the OGE
interaction has to be increased. Compared to nonrelativistic
calculations \cite{Bey} we find smaller quark masses \(m_c\) and
\(m_b\).

It is remarkable that the slope of the confining potential comes out
much larger than in nonrelativistic models, where a typical value is
\(b_c \approx 700\) MeV/fm \cite{Bey}. This is mainly due to the fact
that in nonrelativistic calculations the kinetic energy given by
\(\vec{p}^{\,2}/2m\) is overestimated. In semirelativistic models
based on the relativistic expression \(\sqrt{\vec{p}^{\,2}+m^2}\) (see
e.g.  refs.\cite{GoIsg,GRR}) already higher values \(b_c \approx
900-1000\) MeV/fm have to be used to compensate for the smaller
kinetic energy.  Similar values for \(b_c\) have also been found by
Gara and coworkers within the reduced Salpeter approach \cite{Ga}.
The admixture of the negative energy components within our full
Salpeter approach leads to a further enlargement for \(b_c\).

\subsection{Mass spectra}
\begin{figure}
  \centering
  \leavevmode
  \epsfxsize=0.80\textwidth
  \epsffile{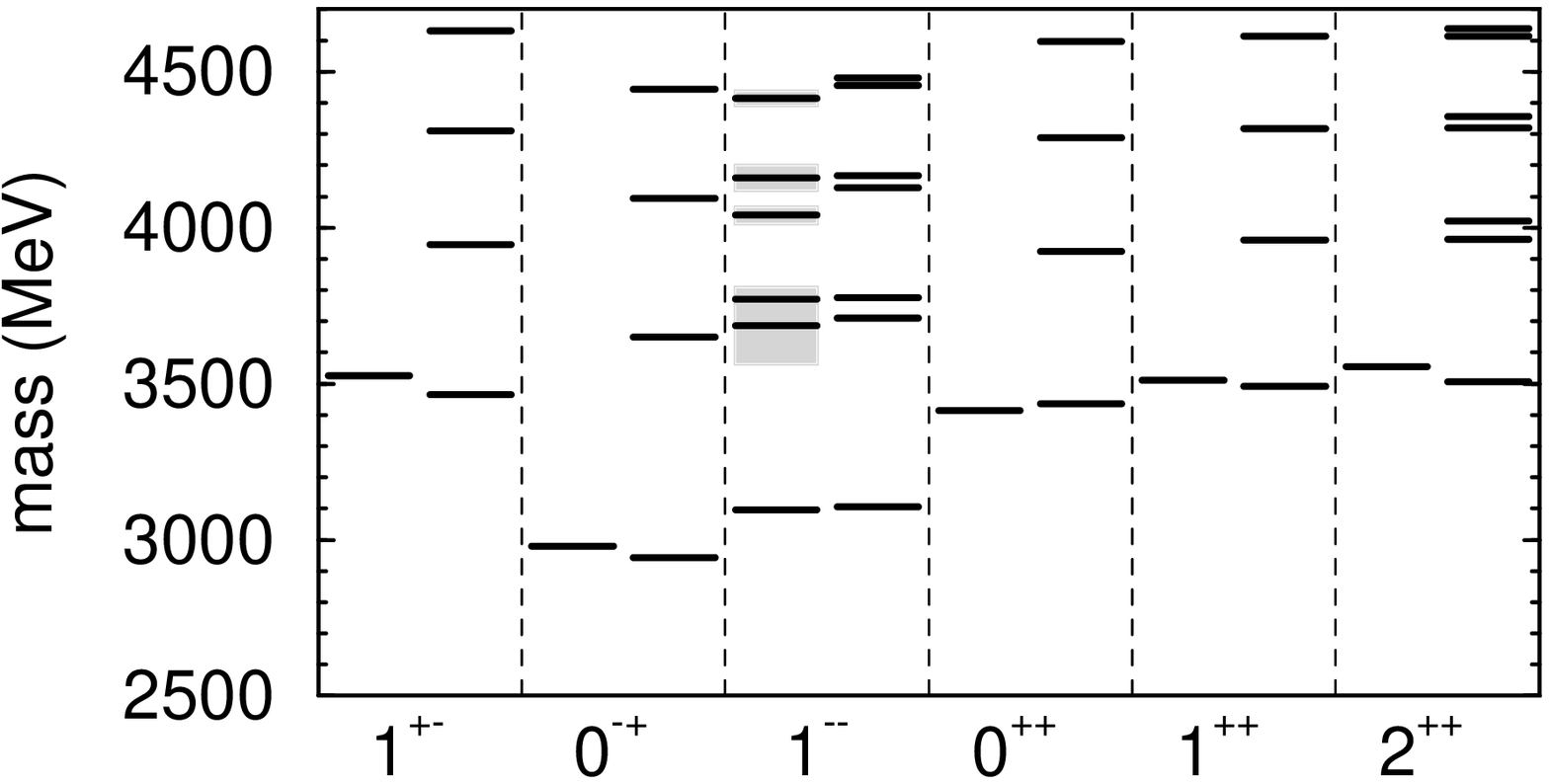}
\vspace{1.5cm}
\caption{Charmonium mass spectrum for a scalar confinement with the
  parameters given in Tab.\protect\ref{ccbbpar}. The left column for
  each meson shows the experimental masses \protect\cite{PDG}, where
  the shaded areas correspond to the full decay widths.}
\label{ccspecscal}
\end{figure}

\begin{figure}
  \vspace{1.5cm} \centering \leavevmode \epsfxsize=0.80\textwidth
  \epsffile{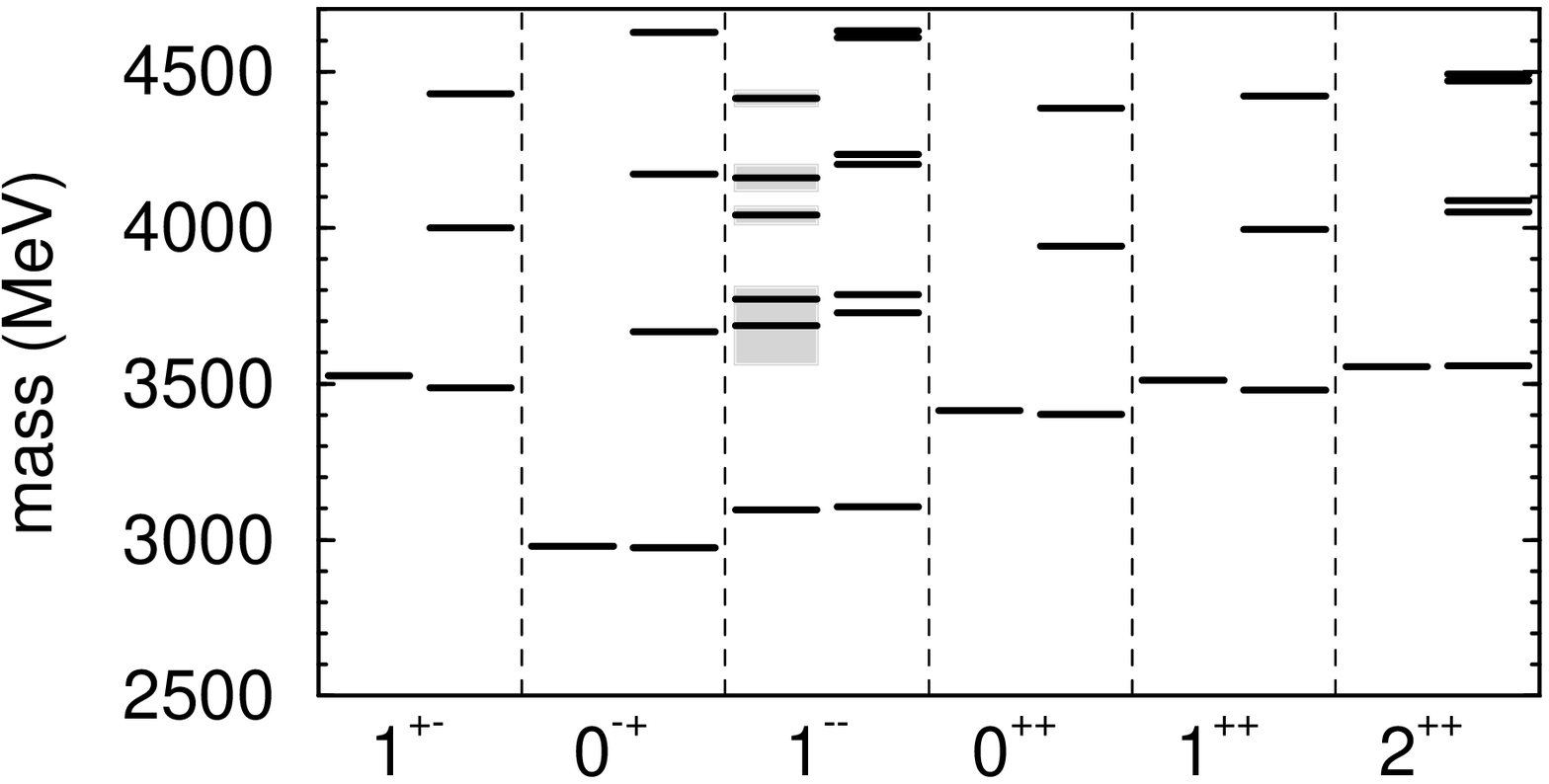} \vspace{1.5cm}
\caption{Same as Fig.\protect\ref{ccspecscal} for a vector confinement.}
\label{ccspecvec}
\end{figure}

\begin{figure}
  \centering \leavevmode \epsfxsize=0.80\textwidth
  \epsffile{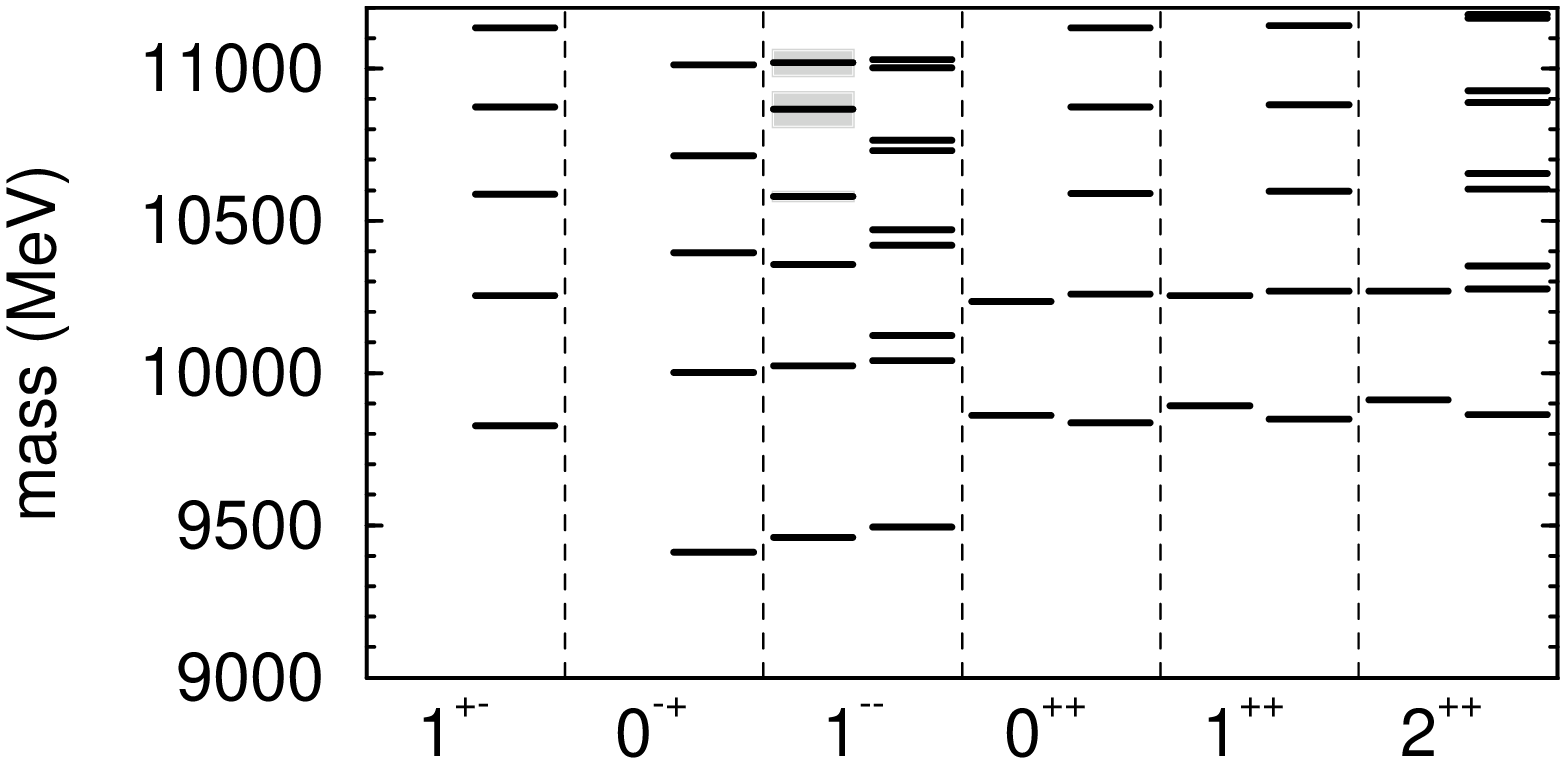} \vspace{1.5cm}
\caption{Bottomonium mass spectrum for a scalar confinement with the
parameters given in Tab.\protect\ref{ccbbpar}. The left column for
each meson shows the experimental masses \protect\cite{PDG},
where the shaded areas
correspond to the full decay widths.}
\label{bbspecscal}
\end{figure}

\begin{figure}
  \vspace{1.5cm} \centering \leavevmode \epsfxsize=0.80\textwidth
  \epsffile{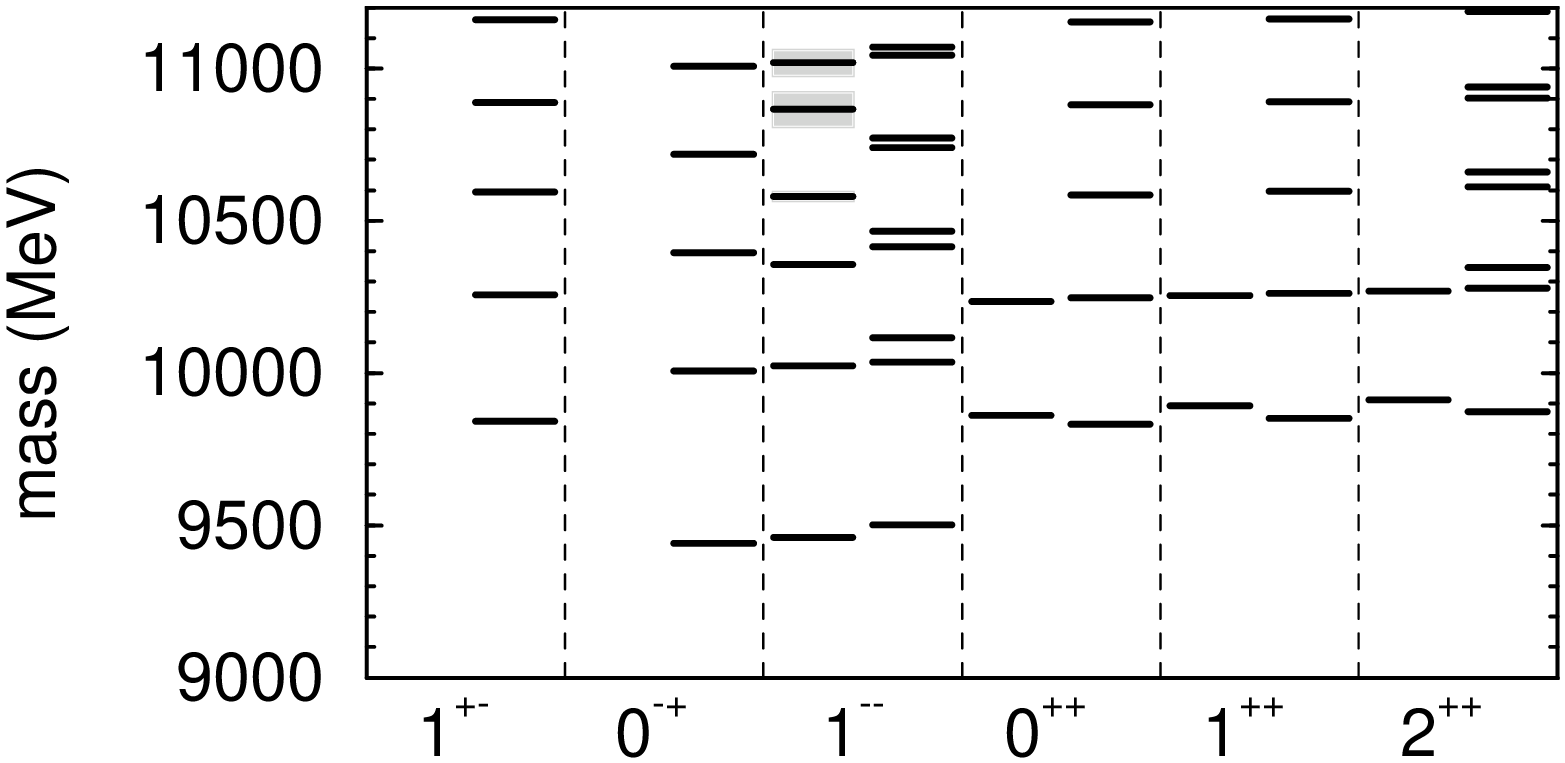} \vspace{1.5cm}
\caption{Same as Fig.\protect\ref{bbspecscal} for a vector confinement.}
\label{bbspecvec}
\end{figure}

The mass spectra of Charmonium are given in
Figs.\ref{ccspecscal},\ref{ccspecvec}, the mass spectra of Bottomonium
are shown in Figs.\ref{bbspecscal},\ref{bbspecvec} for both
confinement spin structures.  The experimental data are usually taken
from the Particle Data Group \cite{PDG}. For the recent measurement of
the mass of the charmonium \({}^1 P_1\) state (\(J^{PC}=1^{+-}\)) in
\(p\bar{p}\) annihilations by the E760 collaboration at Fermilab see
ref.\cite{Arm}.  We find that both confinement spin structures give a
reasonable overall description of the experimental mass spectra. The
spin-spin and spin-orbit splittings are slightly better described for
the vector confinement, whereas the radial excitations of the vector
mesons are slightly better for the scalar confinement. However, we
feel that these differences are not significant enough to decide
wether the Lorentz nature of confinement should be of the scalar or
vector type.  This is in contrast to the nonrelativistic quark model
where a scalar confinement gives the better results.

\begin{figure}
  \centering \leavevmode \epsfxsize=0.55\textwidth
  \epsffile{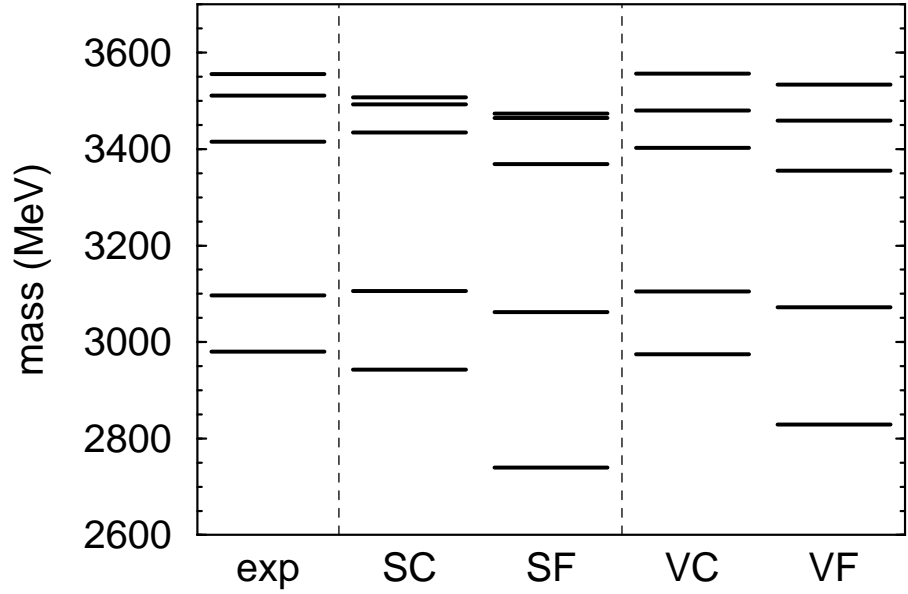} \vspace{1.5cm}
\caption{The charmonium ground states \(\eta_c,\,J/\psi,\,
  \chi_{c0},\,\chi_{c1},\,\chi_{c2}\) (from bottom to top).  The
  coloums correspond (from the left) to the experimental masses, the
  masses obtained with a scalar confinement using the Coulomb (SC) and
  the Feynman gauge (SF), and the same for a vector confinement, i.e.
  (VC) and (VF).  For the Feynman gauge the same parameters have been
  used as for the corresponding Coulomb gauge, see
  Tab.\protect\ref{ccbbpar}.}
\label{Feyngauge}
\end{figure}

Although the description of the mass spectra can be considered quite
satisfactory, there remain some characteristic deviations:
\begin{description}
\item[i)] We find that the binding of the \(\eta_c\) meson tends to be
  quite large. As a consequence it is not possible for a scalar
  confinement to obtain a satisfying simultaneous description of the
  hyperfine splitting \(\eta_c \leftrightarrow J/\psi\) and the fine
  splitting \(\chi_{c0} \leftrightarrow \chi_{c1} \leftrightarrow
  \chi_{c2}\). The problem is less prominent for the vector
  confinement due to the smaller value of \(\alpha_{sat}\) (see
  Fig.\ref{Feyngauge}).
\item[ii)] The large value of the confinement slope \(b_c\) leads to
  an overestimation for the level spacing between the s-wave states of
  the vector mesons, whereas the mass differences of s-waves and
  d-waves is underestimated, especially for higher radial excitations.
\end{description}

To estimate the influence of the gauge chosen for the gluon propagator
we also investigated the Feynman gauge given by
\begin{equation}
\left[V_G^F(\vec{p},\vec{p}\,')\,\Phi(\vec{p}\,')\right]
= {\cal V}_G((\vec{p}-\vec{p}\,')^2)\,
\gamma^{\mu} \Phi(\vec{p}\,') \,\gamma_{\mu} \label{Feyneq}
\end{equation}
As shown in Fig.\ref{Feyngauge} the binding energy of the \(\eta_c\)
meson is overestimated for this gauge. It turns out that it is not
possible to compensate for this effect in a satisfying way by
readjusting the model parameters. The effect of the gauge on the other
states is less important.

It should be noted that due to the large quark masses the
RPA-instability of the Salpeter equation with a scalar confinement as
discussed in refs. \cite{Mue,PaPi} is invisible here for any
accessible number of basis states.  A reasonably small number of basis
states (eleven states have been used in our calculation) thus serves
as a regularization supressing the very high momenta \(|\vec{p}|/m \gg
1\) which lead to the mentioned instability.  We therefore think that
it is legitimate to compare the (quasistable) solutions to the
experimental meson masses.  Note that for light quarks, however, the
instability spoils a reasonable description of light mesons for a
scalar confinement, whereas for a \(\gamma^0 \otimes \gamma^0\)
-vector confinement the solutions remain stable.

\subsection{Decay observables}
As shown in tables.\ref{ccdec} and \ref{E1M1} most decay widths show
only small differences between both confinement spin structures.  The
improvement due to the relativistic treatment is seen most clearly in
the two-photon decay of the \(\eta_c(1S)\) and in the leptonic widths
of the \(\psi(2D)\) and \(\psi(3D)\) (where a larger s-wave admixture,
e.g. due to coupled channel effects, could improve the results).

The leptonic decay widths of the \(c\bar{c}\) s-wave vector mesons are
generally too large by a factor of \(\sim 1.5\) for the \(J/\psi(1S)\)
and more for the higher radial excitations, whereas they are too small
for the \(\Upsilon(1S)\). We were not able to adjust the model
parameters in order to find a better agreement with the experimental
widths, since an increased leptonic width of the \(\Upsilon(1S)\) is
usually connected with an increased \(J/\psi(1S)\) width. Furthermore
the leptonic widths turn out to be quite insensitive to changes of the
parameters which still allow for a reasonable description of the mass
spectra.  The incorporation of the commonly used QCD correction factor
\( (1-16\,\alpha_s/(3\pi)) \) \cite{BGK} does obviously not improve
these results, since the leptonic widths of the \(J/\psi(1S)\) and the
\(\Upsilon(1S)\) would be changed in the same way.

The leptonic widths of the radially excited \(\Upsilon\) states come
out closer to the experimental data. However, the decay widths for
higher radial excitations are too large compared to the widths of the
lower excitations. This is due to the large value of the confinement
slope \(b_c\) which leads to an overestimation of the Salpeter
amplitudes at \(r=0\) for the higher excitations.

For the E1 and M1 transition widths we find some improvement compared
to the nonrelativistic results for the transitions \(\chi_{cJ}(1P)
\rightarrow J/\psi(1S)\,\gamma\), \(\Upsilon(3S) \rightarrow
\chi_{bJ}(2P)\,\gamma\) and \(\Upsilon(2S) \rightarrow
\chi_{bJ}(1P)\,\gamma\). However, for the other transitions the
improvement due to relativistic effects is compensated by the
influence of the large confinement slope \(b_c\) on the Salpeter
amplitudes. Note that transition amplitudes like \(2S \rightarrow 1S\)
are very sensitive to the position of the knot of the 2S amplitude..

\subsection{Comparision with previous results for light mesons}
In two previous papers \cite{Mue,MueForm} we have investigated the
mass spectra, decay widths and electromagnetic form factors of light
mesons within an analogous approach.  For the description of the heavy
quarkonia we have replaced the instanton-induced residual interaction
('t Hooft interaction) applied for the light mesons by the OGE
interaction.  There are two reasons which lead to this different
treatment for light and heavy mesons:
\begin{description}
\item[i)] The similarity of the charmonium and bottomonium mass
  spectra and the mass spectrum of positronium indicates that the OGE
  is a reasonable first approximation of the short-distance
  interaction between heavy quarks. For light quarks, however, the OGE
  leads to degenerate \(\pi\) and \(\eta\) masses in clear
  contradiction with experiment, whereas the 't Hooft interaction
  naturally solves this problem and leads to flavormixing for the
  \(\eta\) and \(\eta'\) mesons.
\item[ii)] The 't Hooft interaction does not give first order
  contributions to the interaction between two charmed or bottom
  quarks because of the flavor antisymmetry of this interaction.
  Effects can only occur via flavor mixing in second order, but the
  large differences in the meson masses suppress such contributions.
  Furthermore there are no experimental indications for other flavors
  contributing significantly to \(c\bar{c}\) and \(b\bar{b}\) mesons.
\end{description}

We found that also for the light mesons a large value of \(b_c=1400\)
MeV/fm (model V2 in ref.\cite{Mue}) had been necessary to describe
higher radial excitations and higher angular momenta \(J>1\). On the
other hand ignoring higher radial excitations and angular momenta
(model V1 in the same reference) enabled a very good description of
the light pseudoscalar and vector meson ground states (i.e. $\pi$,
$\eta$, $K$, $\rho$ etc.) including various decay widths and form
factors (see also ref.\cite{MueForm}).  In this fit a much smaller
value of \(b_c=570\) MeV/fm had to be used which is comparable to
typical values in nonrelativistic calculations.

\section{Summary and conclusion} \label{hIII}
We investigated the heavy quarkonia in the framework of the Salpeter
equation with a linear scalar or vector confinement plus the
one-gluon-exchange interaction. For the mass spectra and the decay
widths we obtained a reasonable overall agreement with the
experimental meson masses for a scalar as well as for a vector
confining kernel. This is in contrast to the nonrelativistic quark
model where a scalar confinement is prefered.

The relativistic framework leads to an improved description especially
for the two-photon decay of the \(\eta_c(1S)\) and the leptonic decays
of the \(\psi(2D)\) and \(\psi(3D)\). Minor improvements are also
found for most E1 transitions. For the other decay widths the
influence of relativistic effects is compensated by the effect of the
large value of the confinement slope \(b_c\).  As a consequence the
masses of the higher radial excitations and the leptonic decays cannot
be described in a satisfying way.

We find that relativistic effects can be important, especially for the
description of certain decays. One would expect that the covariant
formulation applied in the present work should yield a systematic
improvement compared to nonrelativistic calculations.  However, our
results for the heavy quarkonia indicate that this is not the case,
which we blame on the fact that the description of confinement via a
potential is not an adequate concept within a relativistic treatment.

We conclude that despite some success for certain decay widths the
Salpeter approach does not allow for a satisfying description of
higher radial excitations and higher angular momenta.

\begin{appendix}
\section{The OGE potential for small distances} \label{alfasapp}
In this section we will analytically perform the Fourier
transformation of the OGE kernel into coordinate space for \(r \ll
\Lambda^{-1}_{QCD}\) as given by
\begin{eqnarray}
{\cal V}_G^F(r)
&=& \frac{1}{2\pi^2r}\,\int_{|\vec{q}_{low}|}^{|\vec{q}_{high}|}
 |\vec{q}|\,d|\vec{q}|\,\sin(|\vec{q}|r)\,
4\pi\,\frac{4}{3}\,\frac{1}{\vec{q}^{\,2}} \nonumber \\ &\cdot&
\frac{A}{\ln(\vec{q}^{\,2}/\Lambda_{QCD}^2)}
\,\left( 1 - B \,\frac{\ln\,(\ln(\vec{q}^{\,2}/\Lambda_{QCD}^2))}
       {\ln(\vec{q}^{\,2}/\Lambda_{QCD}^2)} \right)
\end{eqnarray}
The cutoff \(|\vec{q}_{low}| \gg \Lambda_{QCD}\) has been introduced
to keep the variable \(|\vec{q}|\) in the high momentum range where
the QCD formula for the running coupling constant is approximately
valid.  The other cutoff \(|\vec{q}_{high}|\) has been introduced for
formal reasons as shown below. It is chosen according to the condition
\(|\vec{q}_{high}|\,r \ll 1/(\Lambda_{QCD}\,r)\).  We basically follow
the way outlined by Lucha et.al. \cite{LSG} who treated the first
order case, i.e. \(B=0\).  Using \(x=|\vec{q}|\,r\) and \(a =
\Lambda_{QCD}\,r \ll 1\) we can write
\begin{equation}
{\cal V}_G^F(r)
= \frac{4}{3r}\, \frac{2\,A}{\pi}\,
\int_{x_{low}}^{x_{high}}
 dx\,\frac{\sin(x)}{x}\,\frac{1}{\ln(x/a)^2}\,
\left( 1 - B \,\frac{\ln\,(\ln(x/a)^2)}
       {\ln(x/a)^2} \right)
\end{equation}
with \(x_{low}=|\vec{q}_{low}|\,r \gg a\) and
\(x_{high}=|\vec{q}_{high}|\,r \ll 1/a\).
Rewrite
\begin{equation}
\frac{1}{\ln(x/a)^2}
= \frac{1}{2\ln a\,(\ln x/\ln a - 1)} =: (*)
\end{equation}
Since \(x \ge x_{low} \gg a\) and \(x < x_{high} \ll 1/a\) we have
\(|\ln x/\ln a| \ll 1\) so that
\begin{equation}
(*) \approx \frac{(-1)}{2\ln a}\,
            \left( 1 + \frac{\ln x}{\ln a} \right)
\end{equation}
For the \(\ln \ln\) -term we further use
\begin{equation}
\ln\,(\ln(x/a)^2)
= \ln\,(2\ln x - 2\ln a)
\approx  \ln(-2\ln a) - \ln x / \ln a
\end{equation}
so that we can write
\begin{eqnarray}
{\cal V}_G^F(r)
&=& \frac{4}{3r}\, \frac{2\,A}{\pi}\,
\int_{x_{low}}^{x_{high}}
 dx\,\frac{\sin(x)}{x}\,
\frac{(-1)}{2\ln a}\, \left( 1 + \frac{\ln x}{\ln a} \right)
\\ &\cdot&
\left( 1 - B \, \frac{(-1)}{2\ln a}\, \left( 1 + \frac{\ln x}{\ln a}
\right)\, \left( \ln(-2\ln a) - \frac{ln x}{\ln a} \right)\,\right)
\nonumber
\end{eqnarray}
It is a good approximation to neglect terms \(\sim (\ln x/\ln a)^2\)
in the following, so that
\begin{eqnarray}
{\cal V}_G^F(r)
&=& \frac{4}{3r}\, \frac{2\,A}{\pi}\,\frac{(-1)}{2\ln a}\,
\int_{x_{low}}^{x_{high}}
 dx\,\frac{\sin(x)}{x}\,
\Bigg[ 1 + \frac{\ln x}{\ln a}
\nonumber \\ &+&
   \left(1+2\,\frac{\ln x}{\ln a}\right)\,B\,
                \frac{\ln (-2\ln a)}{2\ln a}
 - \frac{\ln x}{\ln a}\,\frac{B}{2\ln a} \Bigg]
\end{eqnarray}
In the limit \(r \rightarrow 0\) one has \(a \rightarrow 0\), so that
the limits \(x_{low} \rightarrow 0\) and \(x_{high} \rightarrow
\infty\) can be performed. With the integrals
\begin{equation}
\int_0^{\infty} dx\,\frac{\sin x}{x} = \frac{\pi}{2}
\;\;  ;\;\;\;\;
\int_0^{\infty} dx\,\frac{\sin x}{x}\,\ln x
 = -\frac{\pi}{2}\,\gamma
\end{equation}
where \(\gamma=0.577215\ldots\;\) is the Euler-Mascheroni constant and
with
\begin{equation}
\frac{1}{\ln a}\,\left(1-\frac{\gamma}{\ln a}\right)
\approx \frac{1}{\ln a\,(1+\gamma/\ln a)}
= \frac{1}{\ln (e^{\gamma}\,a)}
\end{equation}
we find
\begin{equation}
{\cal V}_G^F(r)
\approx \frac{4}{3r}\,\Bigg[ \,\frac{-A}{2\ln (e^{\gamma}\,a)}
\,\left( 1 + B\,\frac{\ln (-2\ln a)}{2\ln a} \right)
+\frac{\gamma\,A\,B}{4\,(\ln a)^3}\,\left(
\ln (-2\ln a) - \frac{1}{\ln a} \right)\,\Bigg]
\end{equation}
The term \(\sim 1/(\ln a)^3\) can be neglected to a good approximation
and we finally obtain eq.(\ref{arun}), i.e.
\begin{eqnarray}
  {\cal V}_G^F(r) &\approx& \frac{4}{3}\,\frac{\alpha_s^{run}(r)}{r}
  \;\;\;\mbox{for}\;\; r \ll \Lambda_{QCD}^{-1} \;\;\;\mbox{with} \\
  \alpha_s^{run}(r) &=& \frac{A}{2\,\ln(e^{-\gamma}/a)}\,
  \left[1-B\,\frac{\ln\,(2\,\ln(1/a))}{2\,\ln(1/a)} \right] \nonumber
 \\ &&
\;\;\;\mbox{and}\;\;a=\Lambda_{QCD}\,r
\end{eqnarray}

\end{appendix}

\begin{table}
  \caption{Model parameters for scalar and vector confinement}
  \label{ccbbpar}
\vspace{0.5cm}
  \centering
   \begin{tabular}{ccc}
     \hline
       Parameter                & scalar & vector \\
     \hline
     \(m_c\) [MeV]              & 1507  &  1631  \\
     \(m_b\) [MeV]              & 4857  &  5005  \\
     \(a_c\) [MeV]              & -252  &  -640  \\
     \(b_c\) [MeV/fm]           & 1270  &  1291  \\
     \(\alpha_{sat}\)           & 0.492 &  0.365 \\
     \hline
\end{tabular}
\end{table}

\begin{table}
  \caption{Comparison of experimental and calculated decay
    widths for scalar (S) and vector (V) confinement in \(keV\).  The
    nonrelativistic results (NR) for the \(c\bar{c}\) and \(b\bar{b}\)
    leptonic decay widths are taken from \protect\cite{Bey} (version B
    of the model, nonrelativistic decay formula). The nonrelativistic
    result for \(\eta_c(1S) \rightarrow \gamma\gamma\) has been
    calculated analogously}
  \label{ccdec}
  \centering
   \begin{tabular}{ccccc}
\hline
 decay  &  experimental \cite{PDG} & S & V & NR  \\
     \hline
 \(\Gamma(J/\psi(1S) \rightarrow e^+e^-)\)
      & 5.36 \(\pm\) 0.29  & 8.05 & 9.21 & 12.2 \\
 \(\Gamma(\psi(2S)   \rightarrow e^+e^-)\)
      & 2.14 \(\pm\) 0.21  & 4.30 & 5.87 & 4.63 \\
 \(\Gamma(\psi(2D)   \rightarrow e^+e^-)\)
      & 0.26 \(\pm\) 0.04  & 0.13 & 0.09 & 0.005 \\
 \(\Gamma(\psi(3S)   \rightarrow e^+e^-)\)
      & 0.75 \(\pm\) 0.15  & 3.05 & 4.81 & 3.20 \\
 \(\Gamma(\psi(3D)   \rightarrow e^+e^-)\)
      & 0.77 \(\pm\) 0.23  & 0.23 & 0.14 & 0.01 \\
 \(\Gamma(\psi(4S)   \rightarrow e^+e^-)\)
      & 0.47 \(\pm\) 0.10  & 2.16 & 3.95 & 2.41 \\
\hline
 \(\Gamma(\Upsilon(1S) \rightarrow e^+e^-)\)
      & 1.34 \(\pm\) 0.04  & 0.80 & 0.84 & 1.49 \\
 \(\Gamma(\Upsilon(2S) \rightarrow e^+e^-)\)
      & 0.59 \(\pm\) 0.03  & 0.54 & 0.57 & 0.61 \\
 \(\Gamma(\Upsilon(3S) \rightarrow e^+e^-)\)
      & 0.44 \(\pm\) 0.03  & 0.44 & 0.47 & 0.39 \\
 \(\Gamma(\Upsilon(4S) \rightarrow e^+e^-)\)
      & 0.24 \(\pm\) 0.05  & 0.40 & 0.49 & 0.33 \\
\hline
 \(\Gamma(\eta_c(1S) \rightarrow 2\gamma)\)
      & 6.6 \(\pm\) 2.4    & 4.2  & 3.8 & 19.1\\
\hline
\end{tabular}
\end{table}

\begin{table}
  \caption{Comparison of experimental and calculated E1
    and M1 transition widths for scalar (S) and vector (V) confinement
    given in \(keV\). The estimated error for the calculated widths in
    the Salpeter model is generally smaller than \(10\%\), where the
    number of digits gives a measure of the numerical accuracy.  The
    asterisk indicates that no numerically stable result could be
    obtained.  The nonrelativistic results (NR) are taken from
    \protect\cite{Bey} (reduced version B of the model).}
  \label{E1M1}
  \centering
   \begin{tabular}{ccccc}
\hline
 decay  &  experimental \cite{PDG} & S & V & NR  \\
     \hline
$\psi'(2S) \rightarrow \chi_{c0}(1P)\,\gamma $ &
22.6 $\pm$ 4.5 & 31 & 32 & 19.4 \\
$\psi'(2S) \rightarrow \chi_{c1}(1P)\,\gamma $ &
21.1 $\pm$ 4.2 & 36 & 48 & 34.8 \\
$\psi'(2S) \rightarrow \chi_{c2}(1P)\,\gamma $ &
19.0 $\pm$ 4.0 & 60 & 35 & 29.3 \\
$ \chi_{c0}(1P) \rightarrow J/\psi(1S)\,\gamma $ &
92 $\pm$ 40    & 140 & 119 & 147 \\
$ \chi_{c1}(1P) \rightarrow J/\psi(1S)\,\gamma $ &
240 $\pm$ 40   & 250 & 230 & 287 \\
$\chi_{c2}(1P) \rightarrow J/\psi(1S)\,\gamma $ &
267 $\pm$ 33   & 270 & 347 & 393 \\
\hline
$ \Upsilon(3S) \rightarrow \chi_{b0}(2P)\,\gamma $ &
1.2 $\pm$ 0.4 & 1.4 & 1.5 & 1.00 \\
$ \Upsilon(3S) \rightarrow \chi_{b1}(2P)\,\gamma $ &
2.9 $\pm$ 0.7 & 3.2 & 3.50 & 2.11 \\
$ \Upsilon(3S) \rightarrow \chi_{b2}(2P)\,\gamma $ &
3.1 $\pm$ 0.8 & 3.9 & 4    & 2.59 \\
$ \Upsilon(2S) \rightarrow \chi_{b0}(1P)\,\gamma $ &
1.9 $\pm$ 0.6 & 1.5 & 1.31 & 0.85 \\
$ \Upsilon(2S) \rightarrow \chi_{b1}(1P)\,\gamma $ &
2.9 $\pm$ 0.7 & 2.9 & 2.88 & 1.64 \\
$ \Upsilon(2S) \rightarrow \chi_{b2}(1P)\,\gamma $ &
2.9 $\pm$ 0.7 & 3.8 & 3.40 & 2.00 \\
$ \chi_{b0}(2P) \rightarrow \Upsilon(2S)\,\gamma $ &
              & 13.5 & 13.0 & 13.8 \\
$ \chi_{b1}(2P) \rightarrow \Upsilon(2S)\,\gamma $ &
              & 16 & 15.3 & 15.8 \\
$ \chi_{b2}(2P) \rightarrow \Upsilon(2S)\,\gamma $ &
              & 16.5 & *            & 16.8 \\
$ \chi_{b0}(2P) \rightarrow \Upsilon(1S)\,\gamma $ &
              & 1.45 & 0.95 & 2.52 \\
$ \chi_{b1}(2P) \rightarrow \Upsilon(1S)\,\gamma $ &
              & 2.32 & 2.0 & 6.15 \\
$ \chi_{b2}(2P) \rightarrow \Upsilon(1S)\,\gamma $ &
              & 3.55 & 3.0 & 10.5 \\
$ \chi_{b0}(1P) \rightarrow \Upsilon(1S)\,\gamma $ &
              & 23.2 & 21.5 & 26.2 \\
$ \chi_{b1}(1P) \rightarrow \Upsilon(1S)\,\gamma $ &
              & 26.7 & 25.5 & 30.4 \\
$ \chi_{b2}(1P) \rightarrow \Upsilon(1S)\,\gamma $ &
              & 30.0 & 30.0 & 34.6 \\
\hline
$ \psi'(2S) \rightarrow \eta_c(1S)\,\gamma  $ &
0.7 $\pm$ 0.2 & 6  & 1.3 & 4.47 \\
$ J/\psi(1S) \rightarrow \eta_c(1S)\,\gamma $ &
0.9 $\pm$ 0.3 & 3.35 & 2.66 & 1.21 \\
\hline
\end{tabular}
\end{table}

\end{document}